# Spin-flop magnetoresistance in a collinear antiferromagnetic tunnel junction

Shijie Xu[1,5+], Zhizhong Zhang[1,6+], Farzad Mahfouzi[2+], Yan Huang[1+], Houyi Cheng[1,5+], Bingqian Dai[3], Wenlong Cai[1], Kewen Shi[1], Daoqian Zhu[1], Zongxia Guo[1,4], Caihua Cao[1,4], Yongshan Liu[1,5+], Albert Fert[1], Nicholas Kioussis[2], Kang L. Wang[3], Yue Zhang[1,5]& Weisheng Zhao[1,4,5,6].

[1]Fert Beijing Institute, School of Integrated Circuit Science and Engineering, Beihang University, Beijing, China.
[2]Department of Physics and Astronomy, California State University, Northridge, Los Angeles, CA 91330-8268, USA.
[3]Departments of Electrical and Computer Engineering, Physics and Astronomy, and Material Science and Engineering, University of California, Los Angeles, Los Angeles, CA 90095, USA.
[4]Beihang-Goertek Joint Microelectronics Institute, Qingdao Research Institute, Beihang University, Qingdao, China.
[5]Hefei Innovation Research Institute, Anhui High Reliability Chips Engineering Laboratory, Beihang University, Hefei 230013, China
[6]Zhongfa Aviation Institute, Beihang University Shuanghongqiao Street, Pingyao Town, Yuhang District, Hangzhou 311115, China
[7]These authors contributed equally: Shijie Xu, Zhizhong Zhang, Farzad Mahfouzi, Yan Huang, Houyi Cheng.

Abstract
   Collinear antiferromagnetic (AFM) materials have unique promise of no stray fields, display ultrafast dynamics, and being robust against perturbation filed which motivates the extensive research of antiferromagnetic spintronics. However, the manipulation and detection of antiferromagnetic order remain formidable challenges. Here, we report the electrical detection of colinear antiferromagnetism in all-epitaxial $RuO_2$/MgO/$RuO_2$ three-terminal tunnel junctions (TJ) using spin-flop tunnel anisotropy magnetoresistance (TAMR). We measured a TAMR ratio of around 60% at room temperature, which arises between the parallel and perpendicular configurations of the adjacent collinear AFM state. Furthermore, we carried out angular dependent measurements using this AFM-TJ and showed that the magnitude of anisotropic longitudinal magnetoresistance in the AFM-TJ can be controlled by the direction of magnetic field. We also theoretically found that the colinear antiferromagnetic MTJ may produce a substantially large TAMR ratio as a result of the time-reversal, strong spin orbit coupling (SOC) characteristic of antiferromagnetic $RuO_2$. Our work not only propels antiferromagnetic materials to the forefront of spintronic device innovation but also unveils a novel paradigm for electrically governed antiferromagnetic spintronics, auguring transformative advancements in high-speed, low-energy information devices

**Introduction**

   Antiferromagnetic (AFM) materials have alternating magnetic moments on individual atomic sites [1] This unique property leads to zero stray fields, immunity to external perturbations, and ultrafast spin dynamics [1–3]. Despite its significant potential for ultrafast and ultrahigh-density storage, the outstanding problem is efficiently detecting and manipulating the AFM state[3]. For colinear antiferromagnets (AFMs), which can be stable at small magnetic field, is usually read by imprinting the state on a ferromagnetic tunnel junction (TJ). The detected electrical signal for collinear AFMs due to transverse Hall response or magneto-resistive phenomena is very small [4–10]. Therefore, it is very important to develop a new method to detect collinear antiferromagnetic Néel vector. Recently, the magnetic tunneling effect suitable for detecting the AFM order [11], in which electrical readout enhanced between two magnetic electrodes, have utilized in Magnetic random-access memory (MRAM) and in-memory computing chips [12–14]. Especially for tunneling anisotropic magnetoresistance (TAMR) due to a two-step magnetization, will shows a rich phenomenology that opens new directions in spintronics research.

Recent first-principles theory has predicted a promising route towards spintronics based on spin-neutral currents phenomena, in particular on the AFM tunneling magnetoresistance (ATMR) [15]. The spin-neutral charge current can be controlled by the relative orientation of the Néel vectors of the two colinear AFM electrodes $RuO_2$, resulting in the huge tunneling magnetoresistance (TMR) effect as large as ~500%. In addition, the Néel spin currents emerge in $RuO_2$ with a strong staggered spin-polarization can produce a sizable field-like STT which can deterministic switch the Néel vector of $RuO_2$ [16]. The TMR effect in pure AFM-TJ relies on the orientation of the Néel vector in a single AFM electrode. And the ATMR has hitherto been experimentally demonstrated only in non-collinear AFMs [17, 18]. However, collinear AFMs have unique advantages: they exhibit greater magnetic stability compared to nonlinear AFMs [19, 20]. The parallel or antiparallel alignment of magnetic moments leads to a more regular and stable magnetic structure which have more immunity to external magnetic perturbations. Secondly, the collinear AFMs have stronger exchange interaction between neighboring magnetic moment. In addition, the collinear AFMs have more uniform magnetic properties, making them easier to magnetically engineer and manipulation. Here the $RuO_2$/MgO/ $RuO_2$ tunnel junctions in which large TAMR of the $RuO_2$/MgO interface is combined with efficient rotation of $RuO_2$ moments by the spin-flop effect at the opposite interface of $RuO_2$ with the different thickness. The spin-flop TAMR exceeds 60% at room temperature. We measure all-epitaxial $RuO_2$ AFM-TJ with different magnetic field orientations using magneto-transport measurements. We show that the spin-flop TMR dominates over other magnetic resistance contributions, consistent with theoretical predictions, which is promising for spintronics applications.

**Spin-flop tunneling effect**

Ruthenium dioxide ($RuO_2$) is a typical rutile antiferromagnet with tetragonal crystal structure [10, 21]. Each magnetic atom (Ru) was surrounded by six nonmagnetic oxygen atoms (O). The arrangement of atoms forms a lattice structure with alternating layers of ruthenium and oxygen (Fig. 1a). $RuO_2$ has been reported to host high temperature antiferromagnetism [22, 23] with a Néel temperature $T_N$ >300 K. And the $RuO_2$ also show efficient and broadband terahertz radiation at room temperature due to electrical anisotropy which have been demonstrated in our work [24] [ Fig.1 extend date]. Recently, the RuO$_2$ show the evident anomalous Hall effect, spin splitting effect and tilted spin current in epitaxial single crystal film [10, 21, 25]. However, electrically detecting the AFM order of RuO$_2$ is still an outstanding problem and the readout electrical signal is also very small. The spin-flop effect combined with TMR could bypass those obstacles. The initial state of collinear anti-ferromagnetism is that neighboring magnetic moments align in opposite direction, canceling with each other and resulting in a zero magnetic moment. After increasing the external magnetic field, the alignment of the moments flips toward to a perpendicular direction because of the competition between the Zeeman energy gained from external magnetic field and the exchange energy associated with the AFM exchange interaction between neighboring moments.

To observe the spin-flop tunneling magnetoresistance in collinear AFM materials, it is necessary to fabricate collinear AFM-TJ (Fig. 1b). The all-epitaxial collinear AFM-TJ stacks composed of RuO$_2$ (10 nm)/MgO (2 nm) /RuO$_2$ (20 nm) sandwich structure were grown on the top of oxide SrTiO$_3$ (STO) substrate. The RuO$_2$ (100) grown on the STO (001) was used in recent study due to high temperature collinear Néel order and abundant electronic-magneto transport phenomena of RuO$_2$ [26]. According to the previous work, the MgO based MTJ could have the largest TMR [27, 28] which is currently used in MRAM [14, 29, 30]. The presence of stack films with a smooth and continuous interface between RuO$_2$ and MgO was found by Transmission electron microscopy (TEM) imaging and the measured lattice constant is a =4.5 Å (Fig. 2 a). In addition, the thickness and the surface roughness are fitted from XRR spectra [Fig. 3 extend date] to be close to the measured value of TEM. Furthermore, the Néel vector of RuO$_2$ was found to be perpendicular to the normal direction of the film, allowing us to characterize the films and to measure tunneling conduction in the RuO$_2$/MgO/ RuO$_2$ trilayer film as discussed below. The corresponding selected-area electron diffraction (SAED) studies revealed the top and bottom collinear AFM RuO$_2$ show essentially single crystal feature with the (100) orientation. So, the anomalous Hall can only be zero when the Néel vector is in-plane direction (Fig. 2 extend date). The tunneling layer MgO also show single crystal feature with the (011) orientation with good lattice matching (Fig. 2 b-d) and the all-epitaxial AFM-TJ was prepared on our experiments.

The stacks were then patterned into circular nanopillars with diameters (D) of 3–15 μm on the STO substrate, The Au/Ti top electrode was used for the MTJ readout. The top and bottom RuO$_2$ are required to have two distinctive spin-flop magnetic field so that the parallel and perpendicular configurations of Néel vector arise as a function of magnetic field (Fig.2 e-f). The applied 1uA d.c. current along [001]-axis flowing through the junction, yields a perpendicular measured voltage. Under sweeping magnetic fields along the x direction ([001]- axis) of a device with D = 15 μm, the magnetoresistance loop can be observed at room temperature (Fig.2 g). Because the spin flop field H$_{sp}$ is larger than 2 T, the Néel vector of the RuO$_2$ top layer and RuO$_2$ bottom layer fully stay at in-plane direction at zero fields, corresponding to an initial parallel state of the MTJ with a small tunneling resistance. After switching the magnetic field to 6 T, spin-flop behavior will happen for the 20 nm RuO$_2$ bottom layer, causing to a final perpendicular state of the MTJ with a huge tunneling resistance.

**Angular dependence of Spin-flop tunneling effect**

To further confirm the spin-flop tunneling effect of the all-epitaxial pure AFM TJ, we examined the TAMR in the configuration shown in Fig. 3a. The Néel vector is aligned along the [001]-axis and the current is in the out-plane direction, the θ refers the angle between and the in-plane magnetic field and the Néel vector. Here, the Néel vector tends to be aligned along [001]-axis at initio state. Because both the top and bottom $RuO_2$ are (100) $RuO_2$, both of the top and bottom Néel vector are aligned in-plane direction (Fig. 3. b). The magnetic field dependence of magnetoresistance was clearly observed at different angle θ (Fig. 3. c). For each angle θ, we measured the magnetoresistance by sweeping the in-plane magnetic field within 8 T in which only bottom $RuO_2$ spin moments were switched (explained later). 60% magnetoresistance can be observed only when the magnetic field and Néel vector are parallel, and zero magnetoresistance happened when the magnetic field and Néel vector are perpendicular, proving the in-plane AFM magnetic crystals anisotropy. After rotation of the in-plane magnetic field B from 0 deg to 90 deg, we can find the sin2θ function dependencies between θ and magnetoresistance (Fig. 3 d). Based on single-domain theory, the total AFM magnetic energy $E_m$ of AFM is:

$$E_m = \sum_{i,j} J_{ij} S_i \cdot S_j + \sum_{i,j} D_{ij} S_i \times S_j - K \sum_i S_i \cdot S_i - \sum_i B \cdot S_i$$

Here, is the $J_{ij} S_i \cdot S_j$ is the exchange interaction term, $D_{ij} S_i \times S_j$ is the Dzyaloshinskii-Morriya interaction term, $B \cdot S_j$ is the Zeeman energy under uniform magnetic field. The magnetic anisotropy $E_K = K \sum_i S_i \cdot S_i$ can be explained as: $E_K = K_u sin^2\theta + K_c sin^2 2\theta$, where $K_u$ is the uniaxial anisotropy and Kc is the cubic anisotropy. Therefore, both the uniaxial anisotropy and TAMR have the same sin2θ function dependencies.

To better understand the TAMR mechanism, preliminary results (20x20x1 k-mesh) for the transmission calculation in $RuO_2$[100]/MgO[110]/$RuO_2$[100] tunnel junction versus angle between the Néel orientations of the two $RuO_2$ leads (figure 4.a). Since the interface is non-compensated, there is a conventional TMR effect (i.e. T(theta) proportional to $cos\theta$. The TMR ranges between 10-100% and is relatively insensitive to the SOC (figure 4.b). We do not see any higher order dependence of the transmission versus theta in the absence of SOC (i.e. no spin-flop TMR). When the SOC is turned on, we observe TAMR effect (i.e. T($\theta$) proportional to $cos^2(\theta)$), which is around 10% near the Fermi level and has the potential to reach up to 50%. To our understanding Spin-flop TMR is expected to be non-relativistic and since it is absent in this system, it may best to call it TAMR. There is no obvious relationship between the calculated TMR and the spin-orbit coupling, even if the TMR is larger than the TAMR (figure 4.c). However, the TAMR without considering the spin-orbit coupling is only zero (figure 4.d). Our experiment can match well with the TAMR after considering spin-orbit coupling effect.

**Atomic simulation**

The giant TAMR can also be understood by the following mechanism. we have investigated the magnetic field-induced behavior of RuO$_2$ multilayer membranes. In our simulation, we set the atomic magnetic moment to be 3.5 $\mu_B$, the lattice structure to be Tetragonal lattice. Specifically, our model features a three-layer system, incorporating a central MgO layer sandwiched between two RuO$_2$ layers. For ease of reference, the slender RuO$_2$ layer is denoted as the 'A-layer', and its counterpart as the 'B-layer'. Initiated with Néel vectors and net magnetic moments of both A and B layers parallel to the x-axis, we subjected the system to a horizontally directed magnetic field pointing towards the negative x-axis during the simulations. This model captures not only the influences stemming from the field-generated moments but also the exchange interactions between Ru atoms and the RuO$_2$ lattice, dictating the Ru magnetic anisotropy energy. Experimental data further enlightened us on the magnetic system dynamics under stress; we postulated that interfacial anisotropy undergoes modification due to such stresses. Our comprehensive atomic model of the RuO$_2$ multilayer film captures the magnetic dynamic process triggered by the external magnetic field. As our results elucidate (refer to the accompanying figure 5), while the A-layer RuO$_2$ layer remains static attributed to stress-induced magnetic anisotropy energy the B-layer, because it is thicker, the effect of interfacial stress on the anisotropic properties cannot spread throughout the RuO$_2$ film layer. When influenced by the magnetic field, it undergoes a nearly 90-degree switching. Intriguingly, this behavior mirrors the experimentally observed change in TAMR.

**Conclusions**

The TAMR in all epitaxially collinear antiferromagnet is mainly derived from the spin orbit coupling (SOC). In our experiment, we simultaneously discover the giant TAMR exceed 60% at room temperature and the evident in-plane anisotropy in AFM RuO$_2$. We also clarify the spin flop dynamic by atomic simulation which cause two different AFM interface. The *ab initio* calculation can match with our experiment vert well and explain the TAMR in AFM-TJ. This work proves the giant TAMR in pure collinear AFM. It will deepen the understanding of the tunneling effect in collinear antiferromagnetic materials and facilitates the development of AFM spintronic devices. Our experiments provide a new perspective on tunnel magnetoresistance in structures with two collinear AFM contacts. As we have seen here, the existence of a spin flop effect which need large magnetic field (≥2T). In addition, two different interfacial AFM order lead to the TAMR. We also note that electrical-current-induced writing AFM-TJ remains an exciting challenge for the spintronics device.
.

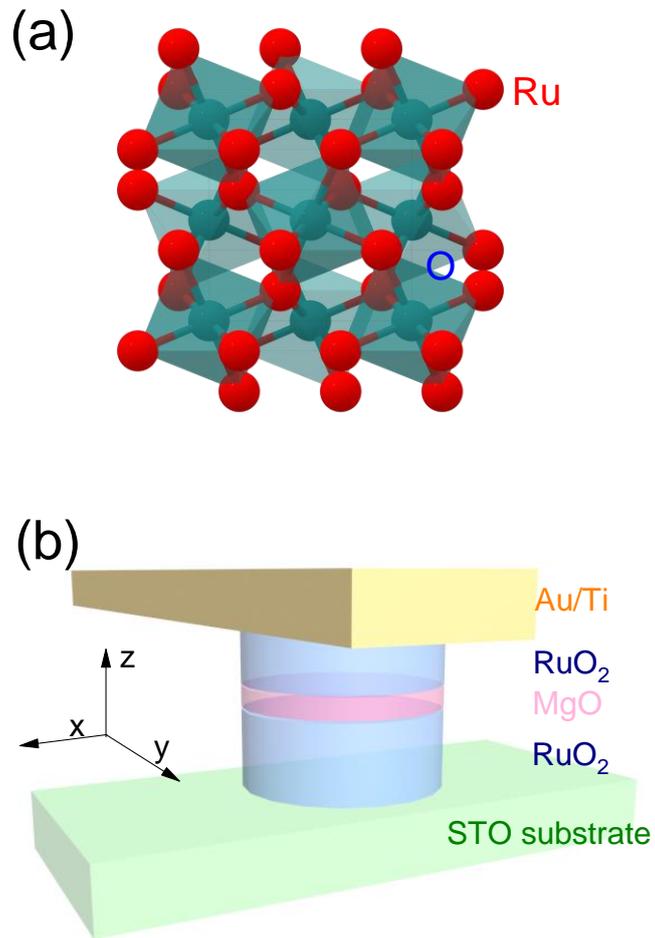

Figure1 a, The atomic structure of RuO$_2$ (the red sphere denotes Ru atom, the blue sphere denote O atom respectively. b, Schematic of the three-terminal MTJ. The bottom RuO$_2$ and top Au/Ti is the writing and detecting channel in the perpendicular direction.

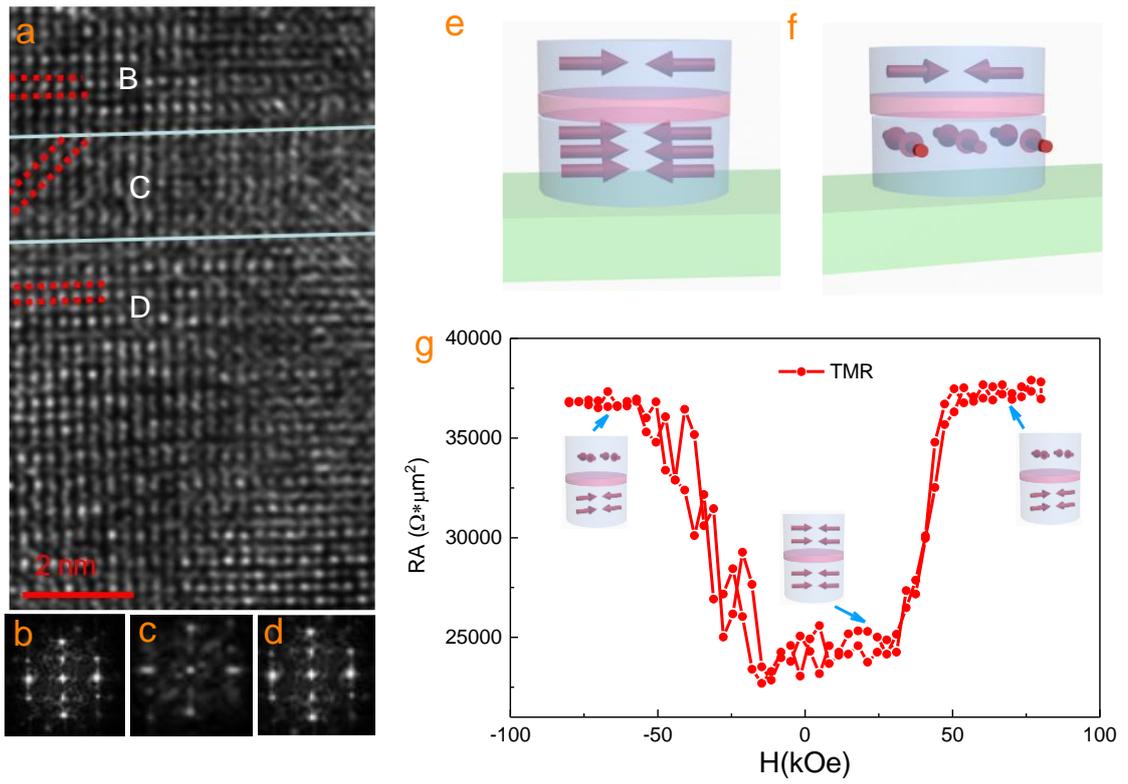

Figure2 a, Transmission electron microscopy(TEM) image of the details of the $RuO_2/MgO/RuO_2$ MTJ. b-d, Selected-area electron diffraction (SAED) patterns of MTJ obtained by TEM. e-f, Schematic of the two states of the bottom $RuO_2$ layer, that is, perpendicular and parallel states of MTJ, respectively. g. MTJ resistance, RA, as a function of external magnetic field,

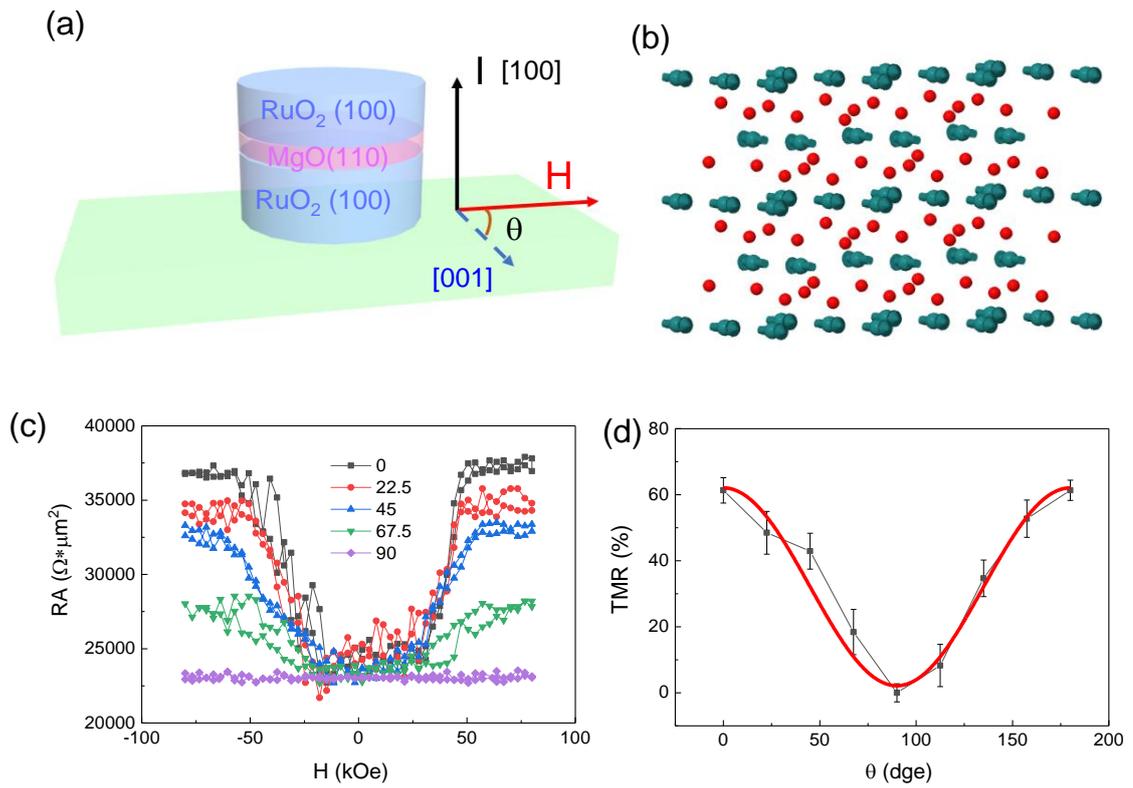

Figure3 a, Schematic of a tunnel junction device. The current is the perpendicular direction. b, Crystal and spin structure of rutile $RuO_2$. Green arrows and red spheres represent Ru and O atoms, respectively. c, MTJ resistance, RA, as a function of external magnetic field at different angle. d. MTJ resistance, RA, as a function of angle at room temperature.

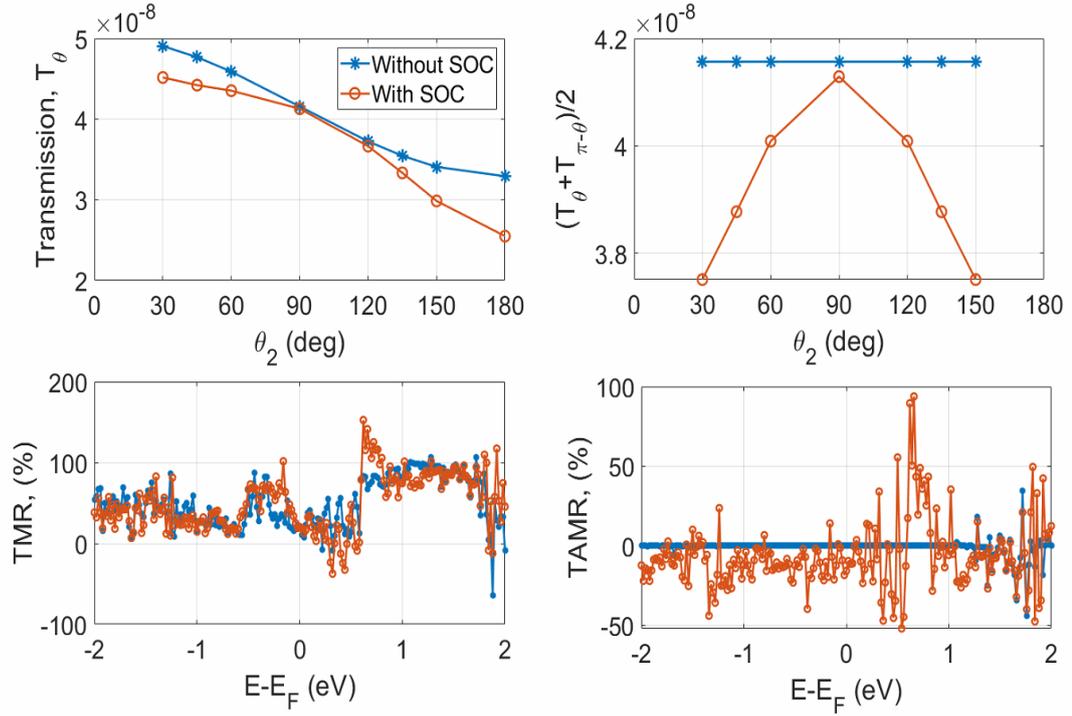

Figure4 a-b, Calculated transmission versus angle in RuO2[100]/MgO[110]/RuO[100] tunnel junction c, Calculated TMR versus Fermi energy. d. Calculated TAMR versus Fermi energy. The blue line denotes calculated result considering spin orbit coupling. The red line denotes calculated result without spin orbit coupling.

Figure 5

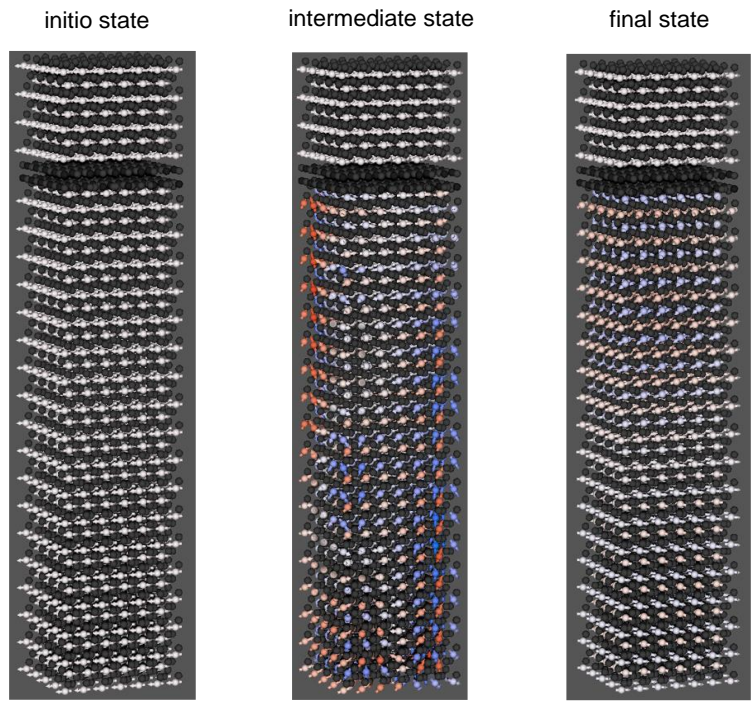

initio state    intermediate state    final state

Figure5, Schematic of RuO$_2$/MgO/RuO$_2$ MTJ in the atomistic simulations. The grey arrow denotes Ru atom.

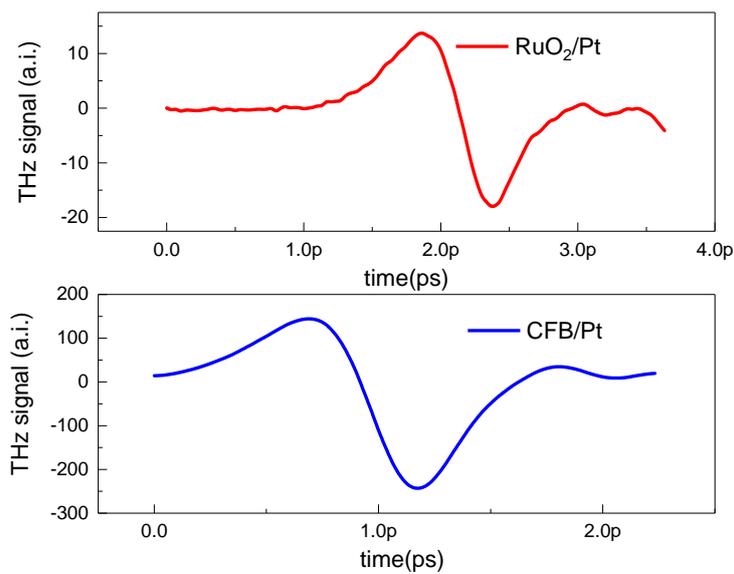

Extend figure 1, Terahertz radiation of RuO$_2$ (101) (8nm)/Pt (4nm).

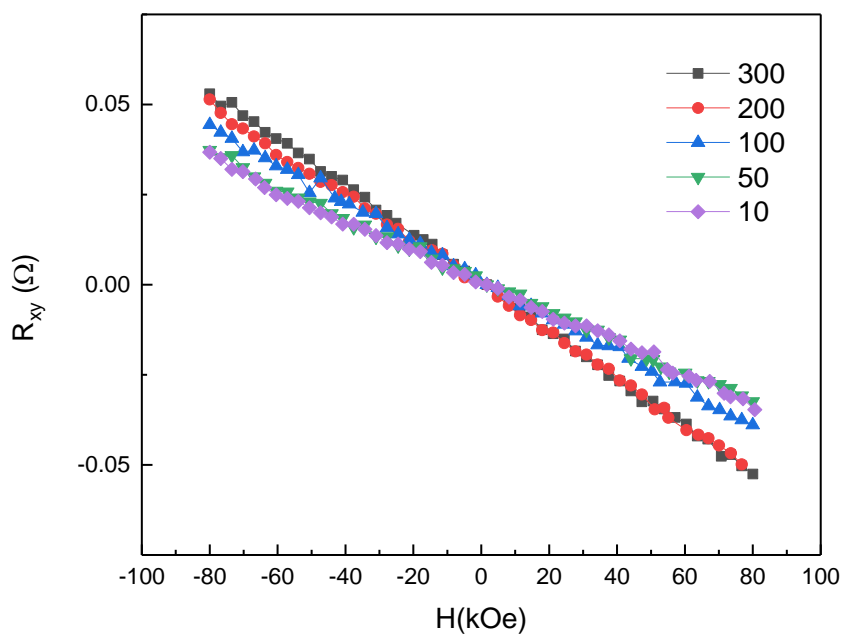

Extend figure 2, Hall effect of RuO$_2$ (100) (20nm).

Extend figure 3

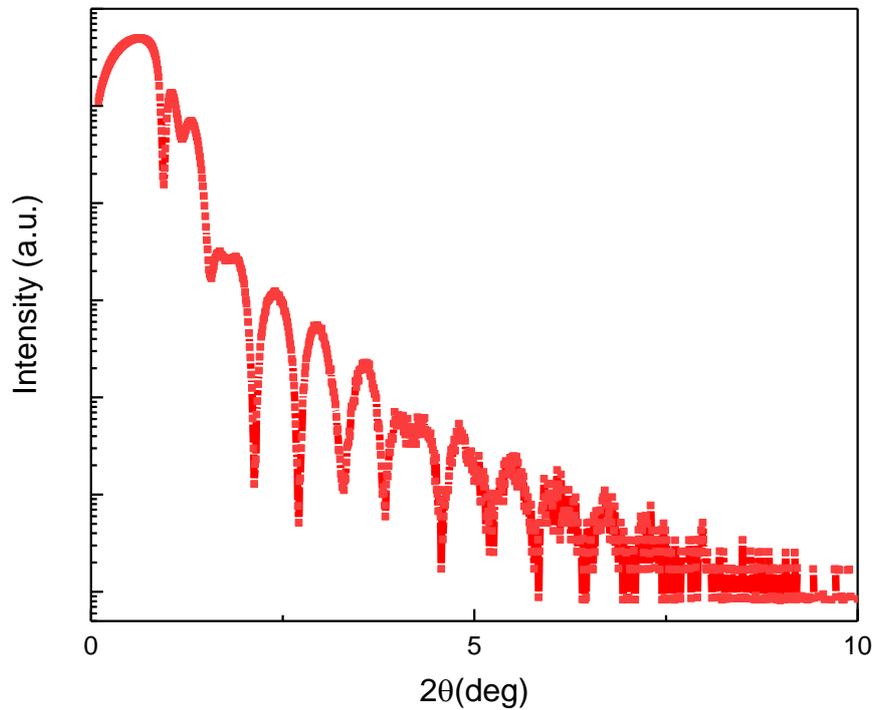

Extend figure 3, X ray reflection (XRR) of $RuO_2$/MgO/$RuO_2$ MTJ (20nm).

# References

[1] V. Baltz, A. Manchon, M. Tsoi, T. Moriyama, T. Ono, and Y. Tserkovnyak, "Antiferromagnetic spintronics," *Rev. Mod. Phys.*, vol. 90, no. 1, p. 15005, 2018, doi: 10.1103/RevModPhys.90.015005.

[2] M. B. Jungfleisch, W. Zhang, and A. Hoffmann, "Perspectives of antiferromagnetic spintronics," *Physics Letters A*, vol. 382, no. 13, pp. 865–871, 2018, doi: 10.1016/j.physleta.2018.01.008.

[3] T. Jungwirth, X. Marti, P. Wadley, and J. Wunderlich, "Antiferromagnetic spintronics," *Nat. Nanotechnol.*, vol. 11, no. 3, pp. 231–241, 2016, doi: 10.1038/nnano.2016.18.

[4]  T. Higo, K. Kondou, T. Nomoto, M. Shiga, S. Sakamoto, X. Chen, D. Nishio-Hamane, R. Arita, Y. Otani, S. Miwa, and S. Nakatsuji, "Perpendicular full switching of chiral antiferromagnetic order by current," *Nature*, vol. 607, no. 7919, pp. 474–479, 2022, doi: 10.1038/s41586-022-04864-1.

[5]  S. DuttaGupta, A. Kurenkov, O. A. Tretiakov, G. Krishnaswamy, G. Sala, V. Krizakova, F. Maccherozzi, S. S. Dhesi, P. Gambardella, S. Fukami, and H. Ohno, "Spin-orbit torque switching of an antiferromagnetic metallic heterostructure," *Nat Commun*, vol. 11, no. 1, p. 5715, 2020, doi: 10.1038/s41467-020-19511-4.

[6]  S. Arpaci, V. Lopez-Dominguez, J. Shi, L. Sánchez-Tejerina, F. Garesci, C. Wang, X. Yan, V. K. Sangwan, M. A. Grayson, M. C. Hersam, G. Finocchio, and P. Khalili Amiri, "Observation of current-induced switching in non-collinear antiferromagnetic IrMn3 by differential voltage measurements," *Nat Commun*, vol. 12, no. 1, p. 3828, 2021, doi: 10.1038/s41467-021-24237-y.

[7]  J. Shi, V. Lopez-Dominguez, F. Garesci, C. Wang, H. Almasi, M. Grayson, G. Finocchio, and P. Khalili Amiri, "Electrical manipulation of the magnetic order in antiferromagnetic PtMn pillars," (in En;en), *Nat Electron*, vol. 3, no. 2, pp. 92–98, 2020, doi: 10.1038/s41928-020-0367-2.

[8]  T. C. Schulthess and W. H. Butler, "Consequences of Spin-Flop Coupling in Exchange Biased Films," *Phys. Rev. Lett.*, vol. 81, no. 20, pp. 4516–4519, 1998, doi: 10.1103/PhysRevLett.81.4516.

[9]  I. S. Jacobs, "Spin-Flopping in MnF2 by High Magnetic Fields," *J. Appl. Phys.*, vol. 32, no. 3, S61-S62, 1961, doi: 10.1063/1.2000500.

[10] Z. Feng, X. Zhou, L. Šmejkal, L. Wu, Z. Zhu, H. Guo, R. González-Hernández, X. Wang, H. Yan, P. Qin, X. Zhang, H. Wu, H. Chen, Z. Meng, L. Liu, Z. Xia, J. Sinova, T. Jungwirth, and Z. Liu, "An anomalous Hall effect in altermagnetic ruthenium dioxide," (in En;en), *Nat Electron*, vol. 5, no. 11, pp. 735–743, 2022, doi: 10.1038/s41928-022-00866-z.

[11] B. G. Park, J. Wunderlich, X. Martí, V. Holý, Y. Kurosaki, M. Yamada, H. Yamamoto, A. Nishide, J. Hayakawa, H. Takahashi, A. B. Shick, and T. Jungwirth, "A spin-valve-like magnetoresistance of an antiferromagnet-based tunnel junction," *Nat. Mater.*, vol. 10, no. 5, pp. 347–351, 2011, doi: 10.1038/nmat2983.

[12] W. Zhao, X. Zhao, B. Zhang, K. Cao, L. Wang, W. Kang, Q. Shi, M. Wang, Y. Zhang, Y. Wang, S. Peng, J.-O. Klein, L. A. de Barros Naviner, and D. Ravelosona, "Failure Analysis in Magnetic Tunnel Junction Nanopillar with Interfacial Perpendicular Magnetic Anisotropy," *Materials (Basel, Switzerland)*, vol. 9, no. 1, 2016, doi: 10.3390/ma9010041.

[13] Z. Guo, J. Yin, Y. Bai, D. Zhu, K. Shi, G. Wang, K. Cao, and W. Zhao, "Spintronics for Energy- Efficient Computing: An Overview and Outlook," *Proc. IEEE*, vol. 109, no. 8, pp. 1398–1417, 2021, doi: 10.1109/JPROC.2021.3084997.

[14] A. Du, D. Zhu, K. Cao, Z. Zhang, Z. Guo, K. Shi, D. Xiong, R. Xiao, W. Cai, J. Yin, S. Lu, C. Zhang, Y. Zhang, S. Luo, A. FERT, and W. Zhao, "Electrical manipulation and detection of antiferromagnetism in magnetic tunnel junctions," (in En;en), *Nat Electron*, vol. 6, no. 6, pp. 425–433, 2023, doi: 10.1038/s41928-023-00975-3.

[15] D.-F. Shao, S.-H. Zhang, M. Li, C.-B. Eom, and E. Y. Tsymbal, "Spin-neutral currents for spintronics," *Nat Commun*, vol. 12, no. 1, p. 7061, 2021, doi: 10.1038/s41467-021-26915-3.

[16] D.-F. Shao, Y.-Y. Jiang, J. Ding, S.-H. Zhang, Z.-A. Wang, R.-C. Xiao, G. Gurung, W. J. Lu, Y. P. Sun, and E. Y. Tsymbal, "Néel Spin Currents in Antiferromagnets," *Phys. Rev. Lett.*, vol. 130, no. 21, 2023, doi: 10.1103/PhysRevLett.130.216702.

[17] P. Qin, H. Yan, X. Wang, H. Chen, Z. Meng, J. Dong, M. Zhu, J. Cai, Z. Feng, X. Zhou, L. Liu, T. Zhang, Z. Zeng, J. Zhang, C. Jiang, and Z. Liu, "Room-temperature magnetoresistance in an all-antiferromagnetic tunnel junction," *Nature*, vol. 613, no. 7944, pp. 485–489, 2023, doi: 10.1038/s41586-022-05461-y.

[18] X. Chen, T. Higo, K. Tanaka, T. Nomoto, H. Tsai, H. Idzuchi, M. Shiga, S. Sakamoto, R. Ando, H. Kosaki, T. Matsuo, D. Nishio-Hamane, R. Arita, S. Miwa, and S. Nakatsuji, "Octupole-driven magnetoresistance in an antiferromagnetic tunnel junction," *Nature*, vol. 613, no. 7944, pp. 490–495, 2023, doi: 10.1038/s41586-022-05463-w.

[19] S. Peng, D. Zhu, W. Li, H. Wu, A. J. Grutter, D. A. Gilbert, J. Lu, D. Xiong, W. Cai, P. Shafer, K. L. Wang, and W. Zhao, "Exchange bias switching in an antiferromagnet/ferromagnet bilayer driven by spin–orbit torque," (in En;en), *Nat Electron*, vol. 3, no. 12, pp. 757–764, 2020, doi: 10.1038/s41928-020-00504-6.

[20] D. Xiong, Y. Jiang, K. Shi, A. Du, Y. Yao, Z. Guo, D. Zhu, K. Cao, S. Peng, W. Cai, D. Zhu, and W. Zhao, "Antiferromagnetic spintronics: An overview and outlook," *Fundamental Research*, 2022, doi: 10.1016/j.fmre.2022.03.016.

[21] H. Bai, Y. C. Zhang, Y. J. Zhou, P. Chen, C. H. Wan, L. Han, W. X. Zhu, S. X. Liang, Y. C. Su, X. F. Han, F. Pan, and C. Song, "Efficient Spin-to-Charge Conversion via Altermagnetic Spin Splitting Effect in Antiferromagnet RuO_{2}," *Physical review letters*, vol. 130, no. 21, p. 216701, 2023, doi: 10.1103/PhysRevLett.130.216701.

[22] Z. H. Zhu, J. Strempfer, R. R. Rao, C. A. Occhialini, J. Pelliciari, Y. Choi, T. Kawaguchi, H. You, J. F. Mitchell, Y. Shao-Horn, and R. Comin, "Anomalous Antiferromagnetism in Metallic RuO_{2} Determined by Resonant X-ray Scattering," *Physical review letters*, vol. 122, no. 1, p. 17202, 2019, doi: 10.1103/PhysRevLett.122.017202.

[23] T. Berlijn, P. C. Snijders, O. Delaire, H.-D. Zhou, T. A. Maier, H.-B. Cao, S.-X. Chi, M. Matsuda, Y. Wang, M. R. Koehler, P. R. C. Kent, and H. H. Weitering, "Itinerant Antiferromagnetism in RuO_{2}," *Physical review letters*, vol. 118, no. 7, p. 77201, 2017, doi: 10.1103/PhysRevLett.118.077201.

[24] S. Zhang, Y. Cui, S. Wang, H. Chen, Y. Liu, W. Qin, T. Guan, C. Tian, Z. Yuan, L. Zhou, Y. Wu, and Z. Tao, "Nonrelativistic and nonmagnetic terahertz-wave generation via ultrafast current control in anisotropic conductive heterostructures," *AP*, vol. 5, no. 05, p. 56006, 2023, doi: 10.1117/1.AP.5.5.056006.

[25] A. Bose, N. J. Schreiber, R. Jain, D.-F. Shao, H. P. Nair, J. Sun, X. S. Zhang, D. A. Muller, E. Y. Tsymbal, D. G. Schlom, and D. C. Ralph, "Tilted spin current generated by the collinear antiferromagnet ruthenium dioxide," (in En;en), *Nat Electron*, vol. 5, no. 5, pp. 267–274, 2022, doi: 10.1038/s41928-022-00744-8.

[26] M. Hasan, R. Dong, H. J. Choi, D. S. Lee, D.-J. Seong, M. B. Pyun, and H. Hwang, "Effect of ruthenium oxide electrode on the resistive switching of Nb-doped strontium titanate," *Appl. Phys. Lett.*, vol. 93, no. 5, 2008, doi: 10.1063/1.2969052.

[27] D. Waldron, V. Timoshevskii, Y. Hu, K. Xia, and H. Guo, "First principles modeling of tunnel magnetoresistance of Fe/MgO/Fe trilayers," *Physical review letters*, vol. 97, no. 22, p. 226802, 2006, doi: 10.1103/PhysRevLett.97.226802.

[28] S. Yuasa, T. Nagahama, A. Fukushima, Y. Suzuki, and K. Ando, "Giant room-temperature magnetoresistance in single-crystal Fe/MgO/Fe magnetic tunnel junctions," *Nature materials*, vol. 3, no. 12, pp. 868–871, 2004, doi: 10.1038/nmat1257.

[29] E. Grimaldi, V. Krizakova, G. Sala, F. Yasin, S. Couet, G. Sankar Kar, K. Garello, and P. Gambardella, "Single-shot dynamics of spin-orbit torque and spin transfer torque switching in three-terminal magnetic tunnel junctions," *Nature Nanotech*, vol. 15, no. 2, pp. 111–117, 2020, doi: 10.1038/s41565-019-0607-7.

[30] H. Honjo, T. V. A. Nguyen, T. Watanabe, T. Nasuno, C. Zhang, T. Tanigawa, S. Miura, H. Inoue, M. Niwa, T. Yoshiduka, Y. Noguchi, M. Yasuhira, A. Tamakoshi, M. Natsui, Y. Ma, H. Koike, Y. Takahashi, K. Furuya, H. Shen, S. Fukami, H. Sato, S. Ikeda, T. Hanyu, H. Ohno, and T. Endoh, "First demonstration of field-free SOT-MRAM with 0.35 ns write speed and 70 thermal stability under 400°C thermal tolerance by canted SOT structure and its advanced patterning/SOT channel technology," in *2019 IEEE International Electron Devices Meeting (IEDM)*, 2019, 28.5.1-28.5.4.